\documentclass[prc, twocolumn]{revtex4-2}

\usepackage{graphicx}
\usepackage{float}
\usepackage[usenames, dvipsnames]{color}
\usepackage[colorlinks]{hyperref}

\usepackage{comment}
\usepackage{amsmath, amssymb, mathrsfs, slashed}
\usepackage{multirow, bm, IEEEtrantools}

\usepackage[normalem]{ulem}
\usepackage{tabularx}

\usepackage[
   top=20mm,
   bottom=20mm,
   left=15mm,
   right=15mm]{geometry}

\usepackage{acro}

\DeclareAcronym{tov}{
  short=TOV,
  long=Tolman-Openheimer-Volkoff,
}
\DeclareAcronym{sm}{
  short=SM,
  long=standard model,
}
\DeclareAcronym{ns}{
  short=NS,
  long=Neutron star,
}
\DeclareAcronym{pns}{
  short=PNS,
  long=proto-neutron star,
}
\DeclareAcronym{hs}{
  short=HS,
  long=hybrid star,
}
\DeclareAcronym{bh}{
  short=BH,
  long=black hole,
}
\DeclareAcronym{qcd}{
  short=QCD,
  long=quantum chromodynamics,
}
\DeclareAcronym{pqcd}{
  short=pQCD,
  long=perturbation quantum chromodynamics,
}
\DeclareAcronym{lqcd}{
  short=lQCD,
  long=lattice quantum chromodynamics,
}
\DeclareAcronym{eos}{
  short=EOS,
  long=equation of state,
}
\DeclareAcronym{nsm}{
  short=NSM,
  long=neutron star matter,
}
\DeclareAcronym{nm}{
  short=NM,
  long=nuclear matter,
}
\DeclareAcronym{ddb}{
  short=DDB,
  long=density depended couplings with Bayesian analysis,
}
\DeclareAcronym{rmf}{
  short=RMF,
  long=relativistic mean-field,
}
\DeclareAcronym{nro}{
  short=NRO,
  long=nonradial oscillation,
}
\DeclareAcronym{ai}{
  short=AI,
  long=artificial intelligence,
}
\DeclareAcronym{gw}{
  short=GW,
  long=gravitational wave,
}
\DeclareAcronym{gr}{
  short=GR,
  long=general relativity,
}
\DeclareAcronym{nicer}{
  short=NICER,
  long=Neutron Star Interior Composition ExploreR,
}
\DeclareAcronym{hp}{
  short=HP,
  long=hadronic phase,
}
\DeclareAcronym{mp}{
  short=MP,
  long=mixed phase,
}
\DeclareAcronym{qp}{
  short=QP,
  long=quark phase,
}
\DeclareAcronym{njl}{
  short=NJL,
  long=Nambu--Jona-Lasinio,
}
\DeclareAcronym{ml}{
  short=ML,
  long=machine learning,
}
\DeclareAcronym{nl}{
  short=NL,
  long=non linear,
}
\DeclareAcronym{pca}{
  short=PCA,
  long=principal component analysis,
}
\DeclareAcronym{qnm}{
  short=QNM,
  long=quasi-normal mode,
}
\DeclareAcronym{hqpt}{
  short=HQPT,
  long=hadron-quark phase transition,
}
\DeclareAcronym{loff}{
  short=LOFF,
  long=Larkin-Ovchinnikov-Fulde-Ferrell,
}
\DeclareAcronym{csc}{
  short=CSC,
  long=color-superconductivity,
}
\DeclareAcronym{cfl}{
  short=CFL,
  long=color-flavor locked,
}
\DeclareAcronym{ur}{
  short=UR,
  long=universal relation,
}

\usepackage{tikz,xcolor,hyperref}

\definecolor{lime}{HTML}{A6CE39}
\DeclareRobustCommand{\orcidicon}{
	\begin{tikzpicture}
	\draw[lime, fill=lime] (0,0) 
	circle [radius=0.16] 
	node[white] {{\fontfamily{qag}\selectfont \tiny ID}};
	\draw[white, fill=white] (-0.0625,0.095) 
	circle [radius=0.007];
	\end{tikzpicture}
	\hspace{-2mm}
}

\usepackage{orcidlink}
\usepackage{academicons}
\newcommand{\orcid}[1]{\href{https://orcid.org/#1}{\textcolor[HTML]{A6CE39}{\orcidicon}}}

\newcommand{\blue}{\color{black}}

\begin{document}

\title{Modification of the universal relation between mass, radius and nonradial $f$-mode oscillation in proto-neutron stars}

\author{Deepak Kumar\orcid{0000-0001-9292-3598}$^{1,2}$} \email{deepak@iiserb.ac.in}
\author{Asit Karan\orcid{0009-0005-3070-2729}$^{1}$} \email{asit22@iiserb.ac.in}
\author{Anshuman Verma\orcid{0000-0003-1103-0742}$^{1}$} \email{anshuman18@iiserb.ac.in}
\author{Hiranmaya Mishra\orcid{0000-0001-8128-1382}$^{2,3}$} \email{hiranmaya@niser.ac.in}
\author{Ritam Mallick\orcid{0000-0003-2943-6388}$^{1}$} \email{mallick@iiserb.ac.in}


\affiliation{$^{1}$Department of Physics, Indian Institute of Science Education and Research, Bhopal, 462 066, India}
\affiliation{$^{2}$Institute of Physics, Sachivalaya Marg, Bhubaneswar 751005, India}
\affiliation{$^3$School of Physical Sciences, National Institute of Science Education and Research, An OCC of Homi Bhabha National Institute, Jatni - 752050, India}

\begin{abstract}
{\blue Neutron stars are usually assumed to be cold; however, in certain dynamical astrophysical scenarios such as newly born neutron stars or binary star mergers, the temperature effects play a non-negligible role. We systematically derive the equation of state at finite-temperature within a relativistic mean-field hadronic model applicable to such proto-neutron stars. The equation of state so derived considerably affects the mass-radius curve, thereby affecting the nonradial quadruple $f$-mode oscillation frequencies.} Temperature effectively makes the equation of state stiffer at relatively low and intermediate densities, thereby making the star less compact and flattening the mass-radius curve. The $f$-mode frequency for low and intermediate-mass neutron stars decreases with temperature and thus should be easier to detect. The universal relation (connecting $f$-mode frequency, mass, and radius) changes nonlinearly with temperature. The parameters defining the universal relation [$\omega M = a(T) \left(\frac{M}{R}\right) + b(T)$] becomes temperature dependent with the coefficients following a parabolic relation with temperature.
\end{abstract}
\maketitle

\section{Introduction}

\ac{ns}s are the densest objects in the universe after the black holes that emerge as remnants of supernovae explosions. As natural astrophysical laboratories, these stellar remnants provide valuable insights into the fundamental physics, astrophysics, and the behavior of matter under extreme densities, pressures, and temperatures. The density in the core can reach up to a few times nuclear saturation density ($\rho_0 = 0.16\ {\rm fm}^{-3}$) and temperature can reach up to a few tens of mega electron-volts at the birth of \ac{ns}. The matter under such extreme conditions is interesting to study, as it gives insights into the properties of strong interactions at a fundamental level \cite{Alford:2019oge}. Indeed, at such extreme conditions, one can expect different phases of normal nuclear matter, such as hadron-quark phase tran-
sition \cite{Pereira:2020cmv, Kumar:2021hzo, Haque:2022dsc, Mallick:2020bdc, Prasad:2017msy}, Larkin-Ovchinnikov-Fulde-Ferrell \cite{Mannarelli:2006fy, Rajagopal:2006ig}, color-flavor locked \cite{Singh:2020bdv, Alford:1998mk, Roupas:2020nua, Chu:2024xpm, Kurkela:2024xfh}, color-superconductivity \cite{Mishra:2022pee, Abhishek:2018xml, Mannarelli:2015bda} or other exotic matter phases \cite{Pal:2024afl, Barbat:2024yvi, Lu:2024kiz, Bhattacharyya:2006vy, Prasad:2019kuz, Prasad:2022dom} including pion and kaon condensed phases \cite{BAYM197429, 10.1143/PTP.80.22} along with usual nuclear matter.

NSs emerge from the supernova explosion, where the inner core of a massive star {(more than 8 $M_{\odot}$ \cite{Farrell:2021xun})} undergoes a rapid collapse because of unbalancing gravitational force, leading to an increase in temperature and density. The gravitational energy released during the collapse results in shock waves that propel the star's outer layers into space and leave behind a compact, extremely hot core, which we call a \ac{pns}. At this point, the electron neutrino attains a finite chemical potential for a while (up to 10--20 s) due to its small mean free path. After that, electron neutrinos escape the core, and again, the matter attains hydrostatic equilibrium, i.e., the inward gravitational force is balanced with the outward degenerate gas pressure. The temperature of the nascent \ac{ns} can reach upto 60 MeV \cite{Fischer:2020xjl, Pons:1998mm, Burrows:1986me}. After its birth, it starts to cool. In a few hundred years, it cools down to the temperature of a few hundred electron-volts \cite{Negreiros:2011aa, Yakovlev:2004iq}.
{Right after their birth, \ac{ns}s known as \ac{pns}s are unsTable and oscillate considerably\cite{Ott_2004, Mueller:2003fs}.} They can have both radial and nonradial oscillations, and the frequency of their oscillations depends on their physical and thermodynamic properties. The study of stellar oscillations, known as asteroseismology \cite{1980tsp..book.....C}, is a very useful technique to understand the inner structure and composition of such compact objects \cite{ Kumar:2021hzo, Kokkotas_1999}. In recent studies, the radial and nonradial oscillations of \ac{ns}s have become very important because they are very sensitive to the matter present in the interior of \ac{ns}s \cite{Arponen:1972zz}. Different studies have found various universal/quasi-universal relations (simple linear relationships) between the nonradial oscillation frequencies, masses, and radii of the cold \ac{ns}s \cite{Kumar:2021hzo, Thakur:2024ijp}. We therefore wonder whether these relations would hold even for the hot \ac{ns}s or if they deviate from them.

One simple way to indirectly detect nonradial oscillation modes is through \ac{gw}s. The nonradial oscillations of \ac{ns}s dump their energy into the \ac{gw}s signal during the inspiral of the binary \cite{1994ApJ...426..688R}. The oscillating \ac{ns}s emit \ac{gw} with different modes of frequencies, such as $g$, $f$, and $p$, etc., depending on the nature of the restoring forces. Therefore, the oscillation frequencies depend strongly on the internal compositions and the various properties of \ac{ns}s. Among them, $g$-modes have the lowest frequencies and are more likely to be detected with third generation detectors or even with present detectors. However, they usually appear when there is a strong temperature or composition gradient in the \ac{ns}. The $p$-mode frequencies are usually relatively high and are not expected to be detected with present detectors. Therefore, in this study, we focus our attention on the $f$-mode frequency, which lies in the 1- to 3- kHz band. The main aim of the study is to analyze how NS properties, along with the $f$-mode frequency and the \ac{ur}, behave when we have a finite-temperature effect in the matter, which is usually the case with proto-neutron stars. This will help us in understanding and constraining the matter inside hot NSs and proto-neutron stars \cite{Glendenning:1997wn, Kumar:2021hzo, Pradhan:2022vdf}

In the present analysis, we study the asteroseismology of \ac{pns}. We present the $f$-mode nonradial oscillations at different stages of \ac{ns} as it cools down and gives the temperature-dependent \ac{ur}s relating the masses, radii, and the $f$-mode oscillation frequencies of \ac{ns}s. The precision in the measurements of macroscopic properties of \ac{ns}, such as mass, radius, tidal deformability, etc., contributes to our understanding of \ac{eos} of nuclear matter under extreme conditions. It is difficult to determine the true nature of matter due to the significant uncertainties in observations (radius in the electromagnetic spectrum) \cite{Nattila:2015jra, Ozel:2016oaf, Steiner:2010fz, Watts:2016uzu}. Asteroseismology can be another tool to probe matter properties in isolated \ac{ns}. While some of the aspects have been explored in the previous studies \cite{Andersson:1996pn, Torres-Forne:2017xhv, Torres-Forne:2018nzj, Ashida:2024nck, Cavan-Piton:2024ayu, Wang:2024dwq} the present investigation gives some new insights on the finite-temperature aspects of \ac{ns}s. It may be mentioned here that the nonradial $g$-mode oscillations at finite-temperatures have been studied using a metamodel in Ref. \cite{Lozano:2022qsm}; on the other hand, we shall study here the nonradial thermal $f$-mode oscillations within the ambit of \ac{rmf} models for nuclear matter at finite-temperature.

We start with the description of \ac{eos} at various temperatures in the \ac{rmf} model with the density-dependent coupling parameters \cite{Typel:2009sy}. We further present the \ac{ur}s at various temperatures. The structure of the article is as follows. In Sec. \ref{nuclear_matter_equation_of_state}, we discuss the EOS of NS matter in \ac{rmf} theory. We give a brief formalism of nonradial oscillation in Sec. \ref{non.radial.oscillation.modes}. In Sec. \ref{results_and_discussion}, we discuss the results of the present work and summarize it in Sec. \ref{summary_and_conclusion}. Throughout the article, we assume $G = 1 = c = \hbar$.

\section{Nuclear matter equation of state} \label{nuclear_matter_equation_of_state}

In this section, we briefly discuss \ac{eos} of nuclear matter relevant for \ac{pns}s. The \ac{eos} for baryonic matter will be derived at finite-temperature using a \ac{rmf} theoretical model solved at a mean-field label with a Lagrangian given as follows \cite{Mishra:2001py, Kumar:2021hzo, Tolos:2016hhl}:
\begin{IEEEeqnarray}{rCl}
\mathcal{L} &=& \sum_{i} \left[\bar{\psi}_i \bigg(i \gamma_{\mu}\partial^{\mu} - m_i +g_{\sigma i}\sigma-g_{\omega i}\gamma_\mu\omega^\mu \right. 
\left. - \frac{1}{2}g_{\rho i} {\bm \rho}_\mu\cdot{\bm \tau} \gamma^\mu \bigg)\psi_i\right] \nonumber \\
&& + \frac{1}{2}\left[\partial_\mu \sigma\partial^\mu\sigma-m_\sigma^2\sigma^2\right] + \frac{1}{2}\left[-\frac{1}{2} \Omega_{\mu\nu}\Omega^{\mu\nu} + m_\omega^2\omega_\mu\omega^\mu\right]  \nonumber \\
&& + \frac{1}{2}\left[-\frac{1}{2}R_{\mu\nu}R^{\mu\nu}-m_\rho^2{\bm \rho}_\mu{\bm \rho}^\mu\right],
\label{lagrangian}
\end{IEEEeqnarray}
where $\psi_i$ represent the baryon with bare mass $m_i$. The interactions between the baryons are governed by the exchange of mesons ($\sigma$ and $\omega$). The scalar meson ($\sigma$) exchange provides attraction interaction, while the vector meson exchange ($\omega$) provides repulsive interaction. We also include an isovector meson ($\rho$) to have isospin asymmetry between baryons. The symbols $\Omega_{\mu\nu}$, and ${\bm R}_{\mu\nu}$ represent the field strength tensors corresponding to the vector and isovector fields $\omega_\mu$, and ${\bm \rho}_\mu$ respectively. The first and the last terms correspond to the Dirac Lagrangian for baryons and leptons, respectively. The rest of the terms correspond to mesons, where we have taken a Klein Gordan Lagrangian density for the scalar $\sigma$ field and a Proca Lagrangian density for the massive vector fields $\omega_\mu$, and ${\bm \rho}_\mu$ fields. The operators ${\bm \tau}$ are the Pauli isospin matrices. The meson couplings are to be considered as density-dependent as \cite{Typel:2018cap, Typel:1999yq} 
\begin{IEEEeqnarray}{rCl}
g_{\sigma i} &=& g_{\sigma 0}\exp(-(x^{a_\sigma}-1)), \\
g_{\omega i} &=& g_{\omega 0}\exp(-(x^{a_\omega}-1)), \\
g_{\rho i} &=& g_{\rho 0}\exp(-a_\rho(x-1)),
\end{IEEEeqnarray}
where $x = \rho/\rho_0$ and $g_{\sigma 0}$, $g_{\omega 0}$, $g_{\rho 0}$, $a_\sigma$, $a_\omega$, and $a_\rho$ are the model parameters and given in the Table \ref{tab:model_parameters}.

From a given Lagrangian, Eq. (\ref{lagrangian}), we find \ac{eos} with the mean-field approximation. In mean-field approximation, we take the meson fields to be classical mean-fields and retain the quantum nature of the baryonic field, i.e., $\langle\sigma\rangle=\sigma_0$, $\langle\omega_\mu\rangle=\omega_0\delta_{\mu 0}$, $\langle\rho_\mu^a\rangle$ = $\delta_{\mu 0}\delta_{a 3}\rho_{03}$. In this approximation, we can define the effective masses and effective chemical potentials for baryons as 
\begin{IEEEeqnarray}{rCl}
M_i &=& m_i - g_{\sigma i} \sigma_0, \label{effective_mass} \\
\tilde \mu_i &=& \mu_i-g_{\omega i}\omega_0-g_{\rho i}I_{3i}\rho_{03}-g_{\phi i}\phi_0-\Sigma^r.
\label{effective_chemical_potential}
\end{IEEEeqnarray}
where the term $\Sigma^r$ appears in Eq. (\ref{effective_chemical_potential}) because of the density-dependent meson couplings \cite{Kumar:2021hzo, Typel:2009sy, Malik:2021nas}. Historically, it is called the ``rearrangement" term, which is given as follows \cite{Typel:1999yq}:
\begin{IEEEeqnarray}{rCl}
\Sigma^{r} &=& \sum_{i=n,p} \left\lbrace -\frac{\partial g_{\sigma}}{\partial \rho_{\rm B}}\sigma_0 \rho^{i}_{s} + \frac{\partial g_{\omega}}{\partial \rho_{\rm B}}\omega_0 \rho^{i} + \frac{\partial g_{\rho}}{\partial \rho_{\rm B}}\rho_3^0 I^{i}_3 \rho^{i}\right\rbrace. \label{rearrangement_term}
\end{IEEEeqnarray}
In the above $\rho_s^i$ and $\rho^i$ are the scalar and vector density of the $i{\rm th}$ baryon and are defined as
\begin{IEEEeqnarray}{rCl}
\rho_s^i &=& -\gamma \int \frac{d{\bm p}}{(2\pi)^3} \frac{M^i}{\sqrt{{\bm p}^2 + M^i{}^2}} \left(1 - f_{-}^i - f_{+}^i \right) \label{rhos1} 
\\
\rho^i &=& \gamma \int \frac{d{\bm p}}{(2\pi)^3} \left( f_{-}^i - f_{+}^i \right), \label{baryon_density}
\end{IEEEeqnarray}
respectively, where $\gamma = 2$ is the spin degeneracy factor, and the total baryon density is defined as 
\begin{IEEEeqnarray}{rCl}
\rho_{\rm B} &=& \sum_{i = n,p} \rho_{\rm i}. \label{total_baryon_density}
\end{IEEEeqnarray}

By using the Lagrange-Euler equations or minimizing the thermal potential corresponding to the Lagrangian [Eq. (\ref{lagrangian})] with respect to the mean-fields, we can find the meson field equations as follows: 
\begin{IEEEeqnarray}{rCl}
m_{\sigma}^2 \sigma_0 &=& \sum_i g_{\sigma i} \rho_s^i, \label{sig}
\\
m_{\omega}^2 \omega_0 &=& \sum_i g_{\omega i} \rho^i, \label{ome}
\\
m_{\rho}^2 \rho_0 &=& \sum_i g_{\rho i} I_3^i \rho^i. \label{rho}
\end{IEEEeqnarray}
Equations (\ref{sig})--(\ref{rho}) are the coupled equations and need to be solved simultaneously. For the sake of completeness, we here present the \ac{eos} which is as follows:
\begin{IEEEeqnarray}{rCl}
p = - \epsilon + \frac{1}{\beta}s + \sum_{i} \mu_i \rho_i, \label{pressure}
\end{IEEEeqnarray}
where $p$, $\epsilon$, $\beta$ and $s$ represent the total pressure, total energy density, inverse of temperature ($\beta$) and total entropy density of the $i{\rm th}$ baryon, respectively. All these quantities are defined as follows: 
\begin{IEEEeqnarray}{rCl}
\epsilon &=& - \gamma \sum_i \int \frac{d^3{\bm p}}{(2\pi)^3} \sqrt{{M_i}^2 + {\bm p}^2}\left( 1 - f^i_{-} - f^i_{+} \right) \nonumber \\
&& + \sum_i\left(g_{\omega i}\omega_0+g_{\phi i}\phi_0+g_{\rho i}\rho_{03}I_{3i}\right)
\rho^i \\ \nonumber
&& + \frac{1}{2}\left( m_\sigma^2\sigma_0^2 + m_\omega^2\omega_0^2 + m_\phi^2\phi_0^2 + m_\rho^2\rho_{03}^2 \right), \label{energy_density}
\\
s &=& - \gamma \sum_i \int \frac{d^3{\bm p}}{(2\pi)^3} \left[ f^i_{-} \ln f^i_{-} + (1 - f^i_{-}) \ln (1 - f^i_{-})\right. \nonumber \\
&& + \left. \left(- \to + \right)\right], \label{entropy_density}
\end{IEEEeqnarray}
where $f^i_{-}$ ($f^i_{+}$) is the Fermi distribution of the $i{\rm th}$ particle (antiparticle) and is defined as 
\begin{IEEEeqnarray}{rCl}
f^i_{\pm} = \frac{1}{1 + \exp [\beta(\sqrt{{M_i}^2+ {\bm p}^2} \mp \tilde{\mu}_i)]}, \label{fermi_distributions}
\end{IEEEeqnarray}
where $M_{i}$ and $\tilde{\mu}_{i}$ are defined in Eqs. (\ref{effective_mass}) and (\ref{effective_chemical_potential}). The matter inside the core is in $\beta$ equilibrium which decides the chemical potentials and the baryon number densities of the constituents of \ac{nsm}. The following equations are the relations for the $\beta$ equilibrium and charge neutrality conditions,
\begin{IEEEeqnarray}{rCl}
{\mu}_i &=& {\mu}_B - q_i (\mu_E - \mu_{\nu}), \label{beta_equalibrium_rmf}
\\
\sum_{i=n,p,l} \rho_i q_i &=& 0, \label{charge_neutrality}
\end{IEEEeqnarray}
where $\mu_B$, $\mu_E$, and $\mu_{\nu}$ are the baryon, electric chemical potentials, and neutrino chemical potentials and $q_i$ is the charge of the $i{\rm th}$ particle. Neutrinos are trapped in the very early stage of proto-neutron star evolution, during which it is assumed that the lepton fraction is kept constant \cite{Burrows:1986me, Malfatti:2019tpg}. The neutrino's chemical potential is then determined
by the lepton fraction \cite{Pons:1998mm}. The neutrino diffusion deleptonises the core on a timescale of a few tens of seconds (10--15 s). After approximately 15s, the \ac{pns} become lepton-poor, but it is still hot. At the end of the 
deleptonization epoch, the star reaches its maximum central temperatures of the order of 50 MeV. At this stage, with vanishing neutrinos chemical potential, the electron chemical potential gets determined by the charge neutrality condition, Eq. (\ref{charge_neutrality}) and (\ref{baryon_density}) \cite{Ferrari:2002ut}. 
The thermally produced neutrino pairs dominate the emission. The neutrino diffusion cools the star with an increase in neutrino mean free path. When the mean free path becomes of the order of stellar radius, the star becomes transparent to neutrinos. During this phase, matter is opaque to neutrinos,
and then as the star cools down, the stellar matter becomes transparent. Thus, the condition  $\mu_{\nu} = 0$ and nonvanishing temperature mimics the ambient condition subsequent to deleptonization \cite{Pons:1998mm}. In this study,  we are taking no neutrinos are trapped and setting $\mu_{\nu} = 0$ for all temperatures considered here. The leptonic contribution to the energy density, Eq. (\ref{energy_density}), and pressure, Eq. (\ref{pressure}) is also added to obtain the final definition of \ac{eos}s.

\begin{table}[h]
\caption{The \ac{rmf} model parameters \cite{Malik:2022zol, Malik:2022ilb, Typel:1999yq, Kumar:2021hzo} and corresponding nuclear saturation properties of nuclear matter at saturation density $\rho_0 = 0.15\ {\rm fm}^{-3}$ and binding energy per baryon comes $\epsilon_0 = -16.10\ {\rm MeV}$, where $N$ in coupling constants denotes neutron and proton only. \label{tab:model_parameters}}
\begin{tabular}{lccc}
\toprule
{Coupling} & {Numerical} & {Nuclear} & {Numerical} \\
{parameters} & {values} & {saturation} & {values} \\
& & {quantities} & {(MeV)} \\
\hline
$g_{\sigma N0}$ & 9.180364 & $K_{0}$ & 231 \\
$g_{\omega N0}$ & 10.981329 & $Q_{0}$ & -109 \\
$g_{\rho N0}$ & 3.826364 & $Z_{0}$ & 1621 \\
$a_\sigma$ & 0.086372 & $J_{\rm sym, 0}$ & 32.19 \\
$a_\omega$ & 0.054065 & $L_{\rm sym, 0}$ & 41.26 \\
$a_\rho$ & 0.509147 & $K_{\rm sym, 0}$ & -116 \\
 & & $Q_{\rm sym, 0}$ & 966 \\
 & & $Z_{\rm sym, 0}$ & -6014 \\
 \hline
\end{tabular}
\end{table}

\begin{table}[h]
\caption{List of \ac{eos}s that are used in the present investigation. These \ac{eos}s are taken from the CompOSE data library \cite{Typel:2013rza, Oertel:2016bki, CompOSECoreTeam:2022ddl}. \label{tab:eos_models}}
\begin{tabular}{lr}
\toprule
{EOS} & {Refs.} \\
\hline
BHB$\Lambda\Phi$ & \cite{Banik:2014qja, Moller:2019jib, Moller:1996uf, Hempel:2009mc} \\
GRDF1 (DD2)      & \cite{Typel:2018wmm,Typel:2009sy,Typel:2013zna} \\
HSB              & \cite{Hempel:2009mc,TOKI1995c357,2005PThPh.113..785G} \\
HSB SF           & \cite{Hempel:2009mc,Hempel:2011mk,Moller:1996uf} \\
STOS (TM1)       & \cite{Steiner:2012rk, Hempel:2011mk} \\
VE (RNF)         & \cite{Togashi:2017mjp,Shen:1998gq} \\
SHO (FSU2)       & \cite{Hempel:2009mc,Typel:2009sy, Moller:1996uf} \\
\hline
\end{tabular}
\end{table}

{
\section{Neutron star and its nonradial oscillation \label{non.radial.oscillation.modes}}
The general static, spherically symmetric metric that describes the geometry of a static \ac{ns}, can be expressed as
\begin{IEEEeqnarray}{rCl}
ds^2 &=& e^{2\nu(r)} dt^2-e^{2\lambda(r)} dr^2-r^2 (d\theta^2+\sin^2\theta d\phi^2), \label{metric}
\end{IEEEeqnarray}
where, $\nu(r)$ and $\lambda(r)$ are the metric functions. It is convenient to define the mass function, $m(r)$, in terms of $\lambda$(r) as 
\begin{equation}
e^{2\lambda(r)} = \left(1-\frac{2m(r)}{r}\right)^{-1}.
\end{equation}
Starting from the line element in Eq. (\ref{metric}), using Einstein-field equation for the metric, one can derive the equations that govern the structure of spherically symmetric compact objects, known as the {\blue \ac{tov}} equations \cite{PhysRev.55.374, PhysRev.55.364}, as
\begin{IEEEeqnarray}{rCl}
\frac{dp(r)}{dr} &=& -\left(\epsilon(r) +p(r) \right)\frac{d\nu }{dr}, \label{tov.pressure}
\\
\frac{dm(r)}{dr} &=& 4\pi r^2 \epsilon (r), \label{tov.mass}
\end{IEEEeqnarray}
\begin{IEEEeqnarray}{rCl}
\frac{d\nu(r)}{dr} &=& \frac{m(r)+4 \pi r^3p(r)}{r(r-2m(r))}. \label{tov.phi}
\end{IEEEeqnarray}

In the above set of equations, $\epsilon(r)$ and $p(r)$ represent the energy density and pressure, respectively. The function  $m(r)$ denotes the mass enclosed within a radius $r$ of the compact star. The boundary conditions $m(r=0) = 0$, $p(r=0)=p_c$, and $p(r=R) = 0$, where $p_c$ is the central pressure, lead to an equilibrium configuration when combined with the \ac{eos} of neutron star matter. This process determines the radius $R$ and the total mass $M = m(R)$ of the neutron star for a given central pressure $p_c$ or central energy density $\epsilon_c$. By varying $\epsilon_c$, one can construct the mass-radius (M-R) curve.

Using the Einstein-field equations and the conservation of baryon number, the theory for the \ac{nro}s of \ac{ns}s was developed in Ref. \cite{1967ApJ...149..591T}. The perturbation of the fluid within the star is described by the Lagrangian displacement vector $\xi^{\alpha}$, which can be expressed in terms of the perturbing functions $Q(r,t)$ and $Z(r,t)$ as
\begin{IEEEeqnarray}{rCl}
\xi^i = \left(e^{-\lambda(r)}Q(r,t),\ -Z(r,t)\partial_{\theta},\ 0 \right) r^{-2}P_{l}(\cos \theta).
\end{IEEEeqnarray}

We assume a harmonic time dependence for the perturbation functions $Q(r,t)$ and $Z(r,t)$, which are proportional to $e^{-i\omega t}$, where ``$\omega$" represents the oscillation frequency. Furthermore, we do not consider toroidal deformations in this analysis. Under the Cowling approximation \cite{Kumar:2021hzo}, the perturbation functions satisfy, in general, a set of first-order coupled differential equations,
\begin{IEEEeqnarray}{rCl}
Q' - \frac{1}{c_e^2}\left[\omega^2 r^2e^{\lambda-2\nu}Z+\nu' Q\right]+l(l+1)e^\lambda Z &=& 0, \nonumber \\
\label{qprime}
\\
Z' - 2\nu' Z+e^\lambda \frac{Q}{r^2}-\nu'\left(\frac{1}{c_e^2}-\frac{1}{c_s^2}\right)\left(Z+\nu'e^{-\lambda+2\nu}\frac{Q}{\omega^2r^2}\right) &=& 0 \nonumber \\ \label{zprime_chap1},
\end{IEEEeqnarray}
where the prime denotes differentiation with respect to the radial coordinate. For a comprehensive derivation of Eqs. (\ref{qprime}) and (\ref{zprime_chap1}), we refer to \cite{Kumar:2021hzo}. In these equations, $c_e^2 = dp/d\epsilon = p^\prime/\epsilon^\prime$ represents the square of the equilibrium speed of sound, which is determined from the derivative of the \ac{eos} in $\beta$ equilibrium. On the other hand, assuming that weak interaction timescales are much longer than the timescales of \ac{nro}s, the adiabatic sound speed is given by $c_s^2 = \left(\partial p/\partial \epsilon \right)_{y_i,s}$. The prefactor of the last term on the left-hand side of Eq. (\ref{zprime_chap1}) is proportional to the relativistic {\blue Brunt-V\"{a}is\"{a}la} frequency \cite{McDermott:1983}, which is responsible for gravity-mode ($g$-mode) oscillations \cite{Kumar:2021hzo}. {Since we are considering $npe$ matter and no phase transition is taken here, both the sound speeds are same. In the present study, this term is zero. We focus exclusively on fundamental ($f$-mode) oscillations.}

The coupled first-order differential equations for $Q(r)$ and $Z(r)$, given by Eqs. (\ref{qprime}) and (\ref{zprime_chap1}), must be solved with appropriate boundary conditions at both the center and the surface of the star. Near the center of compact stars, the behavior of the functions $Q(r)$ and $Z(r)$ is given by \cite{Sotani:2010mx}:
\begin{eqnarray}
Q(r) = Cr^{l+1} \quad \mathrm{and} \quad Z(r)=-Cr^l/l, \label{intital.conditions.of.w.and.v}
\end{eqnarray}
where $C$ is an arbitrary constant and $l$ is the order of the oscillation. The other boundary condition is the vanishing of the Lagrangian perturbation pressure, i.e., $\Delta p=0$. The vanishing of $\Delta p$ at surface leads to the boundary condition \cite{Kumar:2021hzo} 
\begin{IEEEeqnarray}{rCl}
\left[\omega^2 r^2e^{\lambda-2\nu}Z+\nu' Q\right]_{r=R} = 0. \label{surface_conditions}
\end{IEEEeqnarray}

For a given central pressure, we first solve the \ac{tov} equations, Eqs. (\ref{tov.pressure}) and (\ref{tov.mass}), to obtain the profiles of the unperturbed metric functions $\nu(r)$ and $\lambda(r)$, as well as the mass function $m(r)$ as a function of radial distance from the center of the star. Next, for a given frequency, we solve the pulsation equations, Eqs. (\ref{qprime}) and (\ref{zprime_chap1}). These solutions are then substituted into the left-hand side of Eq. (\ref{surface_conditions}). The value of $\omega$ is varied iteratively to ensure that the boundary condition given by Eq. (\ref{surface_conditions}) is satisfied. This procedure determines the oscillation frequency as a function of the mass and radius of the star. It is important to note that multiple solutions for $\omega$ may satisfy the boundary condition for different initial trial values of $\omega$. Each of these solutions corresponds to a distinct nonradial oscillation mode of the compact star.
}

\section{Results and discussion} \label{results_and_discussion}
One of the key elements of \ac{pns} study is the temperature effects on the \ac{eos} for the NS matter which in turn, affects \ac{ns} structure and the $f$-mode oscillation mode and its corresponding \ac{ur}s.To obtain the temperature-dependent \ac{eos}, our numerical procedure involves a solution of the self-consistent equations for the meson fields [Eq.(\ref{sig}), Eq.(\ref{ome}) and Eq.(\ref{rho})] and the scalar and the baryon number densities for fixed values of temperature and 
and baryon density. As mentioned in Sec. \ref{nuclear_matter_equation_of_state}, we shall take neutrino chemical potential to be zero mimicking
the ambient condition subsequent to deleptonisation.

In the present work, we use the \ac{rmf} theory to get a temperature-dependent \ac{eos}. We also compare our result with other standard finite-temperature \ac{eos}s given in Table \ref{tab:eos_models}. Here the coupling parameters of the theory are density dependent and are given in the Table \ref{tab:model_parameters}. We find nuclear saturation density is $\rho_0 = 0.15\ {\rm fm}^{-3}$ and the binding energy per nucleon is turns out to be $\epsilon_{0} = -16.10\ {\rm MeV}$. At nuclear saturation density, the other nuclear saturation properties are given in Table \ref{tab:model_parameters}. 

\begin{figure}[htbp]
    \centering
    \includegraphics[scale=0.45]{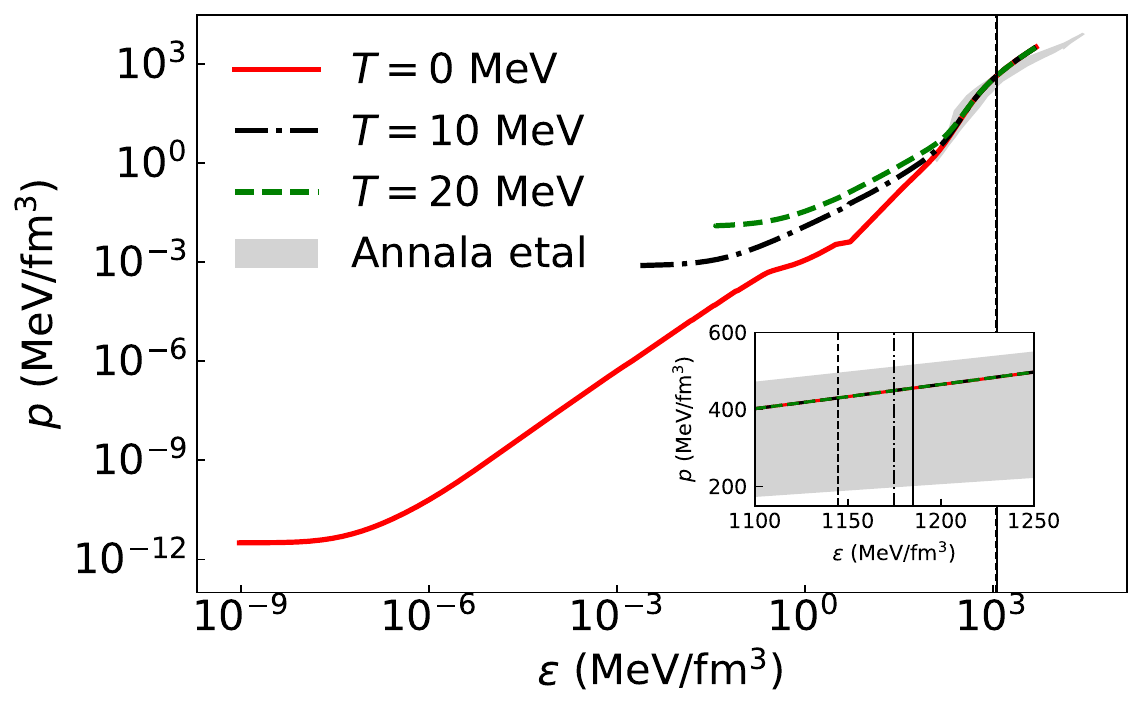}
    \caption{\blue $\beta$-equilibrated \ac{eos} for various temperatures within \ac{rmf} model for nuclear matter. The model parameters are listed in Table \ref{tab:model_parameters}. For lower energy densities, the \ac{eos} is taken from the \ac{eos} for the crustal matter given by Hempel {\it et al.} in Ref. \cite{Hempel:2011mk}. The gray band at higher densities corresponds to the \ac{eos} given by Annala {\it et al.} in Ref. \cite{Annala:2017llu}. In the inset,  the \ac{eos} is shown in a linear scale for the energy densities for the sake of clarity. The vertical lines refer to the central energy densities corresponding to the maximum mass \ac{ns}s for a given temperature.}
    \label{fig:equation_of_state}
\end{figure}
In Fig. \ref{fig:equation_of_state}, we display the \ac{eos} of nuclear ($npe$) matter for various temperatures in the $\beta$-equilibrium and neutrino-free cases. The black dot-dashed (green dashed) curve represents \ac{eos} at temperature $T = 10\ (20)\ {\rm MeV}$ while the solid red curve represents the same in the zero-temperature limit. For lower densities we have adopted here the \ac{eos} of the crust as given by Hempel {\it et al.} \cite{Hempel:2011mk} by matching their thermodynamic potential with that obtained for the \ac{rmf} model for a given density and temperature. The gray band in Fig. \ref{fig:equation_of_state} at higher densities corresponds to the \ac{eos} obtained by Annala {\it et al.} \cite{Annala:2017llu} obtained by interpolating results of chiral effective theory at lower densities and \ac{pqcd} at asymptotically high density. It may be mentioned here that the \ac{eos}, in the zero-temperature limit, satisfies the saturation properties given in Table \ref{tab:model_parameters}. At intermediate densities, it is consistent with the region as given in Ref. \cite{Annala:2017llu}. The \ac{eos} obtained from the \ac{rmf} model deviate from the results of Annala {\it et al.}  Ref. \cite{Annala:2017llu} at larger densities as we have not taken the constraints form \ac{pqcd} which is valid at asymptotically high densities. However, for the \ac{eos} taken here, the central energy density corresponding to the maximum mass \ac{ns} is within the limiting band given in Ref. \cite{Annala:2017llu}. In the inset of Fig. \ref{fig:equation_of_state}, we have shown the central energy densities for the maximum mass \ac{ns}s by the vertical lines. As may be observed, the central energy density corresponding to the maximum mass \ac{ns}s decreases with temperature. Such an observation was also made in Ref. \cite{Sun:2024vdt} for the \ac{rmf} models. {\blue It may further be observed that the temperature effects on the \ac{eos} are prominent at lower densities. At higher densities the baryon chemical potential sets the scale for the \ac{eos} as the temperatures considered are much too small to affect the \ac{eos} at large densities. It is also seen from the figure that the \ac{eos} stiffens with temperature at relatively low densities; however, they are unaffected in the intermediate density range (the density range relevant to the core of NSs).} 

\begin{figure*}[htbp]
    \centering
    \includegraphics[scale=0.45]{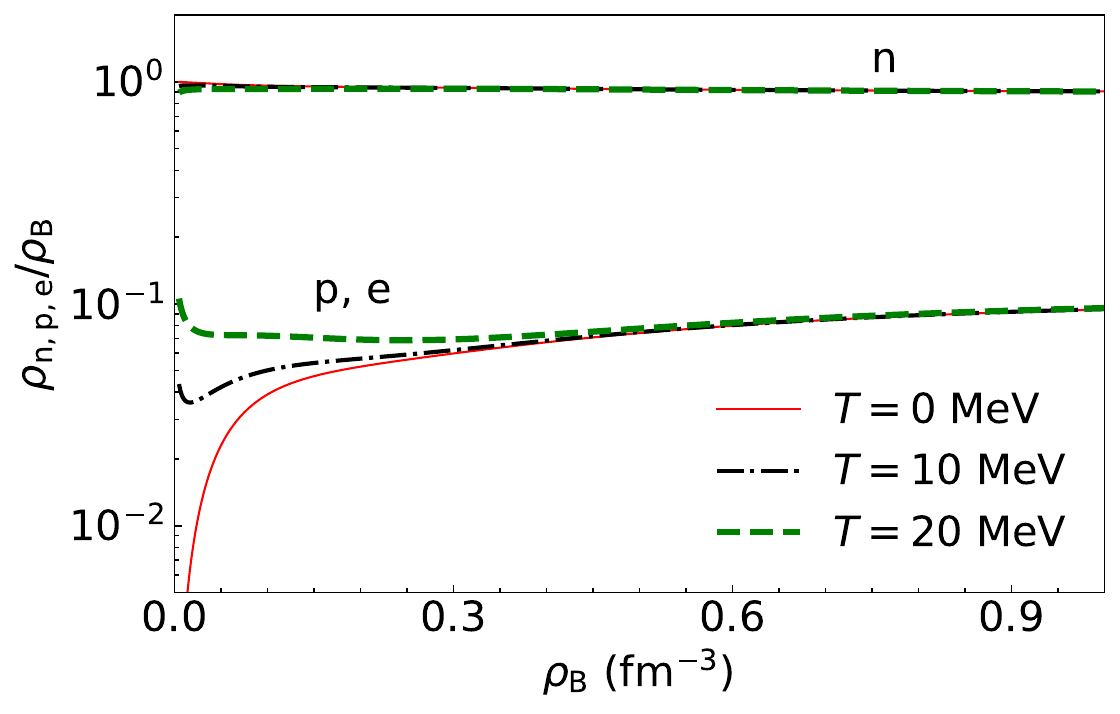}
    \includegraphics[scale=0.45]{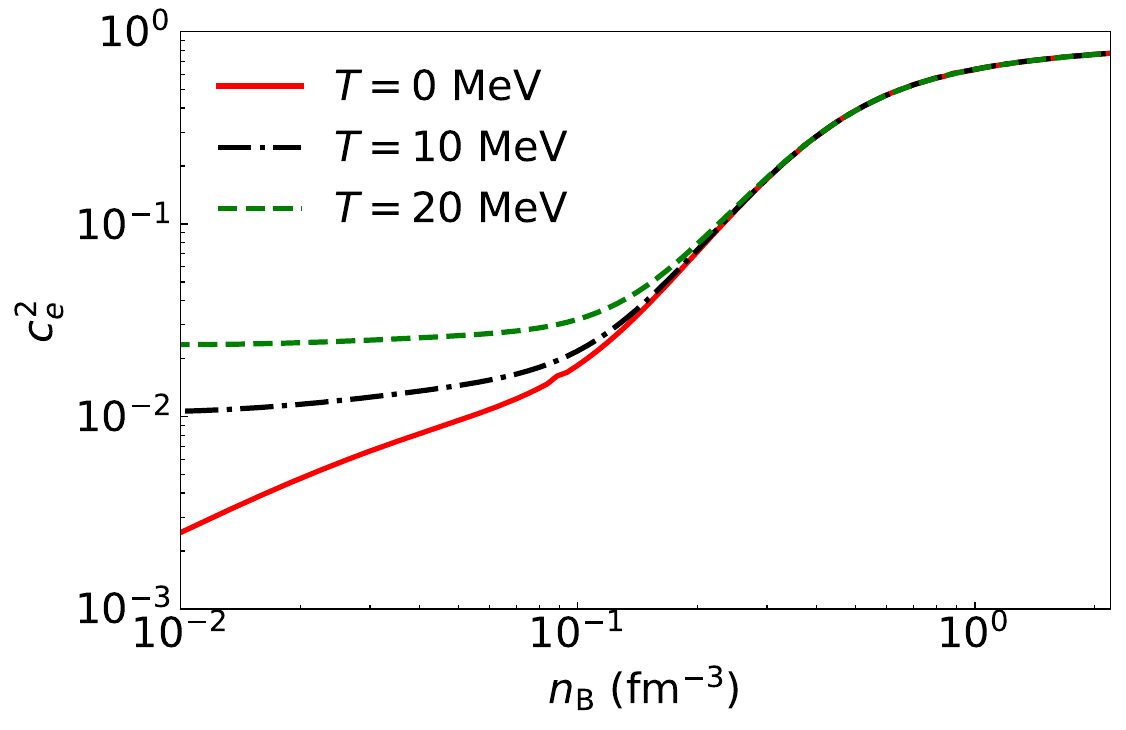}
    \caption{\blue The particle fraction (left) and square of the isothermal speed of sound (right) variations at different temperatures ($T = 0,10,20$\ MeV) as a function of baryon number density. The speed of sound as a function of baryon density is calculated as $c_e^2 = {dp}/{d\epsilon}$ with the $\beta$-equilibrated equation of state. The crust, we are choosing here, is taken from the Hempel {\it et al.} \ac{eos} \cite{Hempel:2011mk}.}
    \label{fig:speed_of_sound_and_particle_densities}
\end{figure*}
In Fig. \ref{fig:speed_of_sound_and_particle_densities} (left), we show the particle fractions as a function of total baryon density for $\beta$-equilibrated, charge neutral and neutrino transparent $npe$ matter of \ac{ns} at various temperatures, $T = 0\ {\rm MeV}$ (red solid curve), and $T = 10\ {\rm MeV}$ (black dot-dashed curve), $T = 20\ {\rm MeV}$ (green dashed curve). The charge neutrality condition forces the equality of proton and electron number densities. As the temperature is increased, there can be nonzero densities of proton (electron), which can be thermally produced, unlike the case of zero-temperature at low baryon densities. This can have important ramifications for hot \ac{ns} matter relevant for \ac{ns} merger oscillations \cite{Sedrakian:2021qjw}. In Fig. \ref{fig:speed_of_sound_and_particle_densities} (right), we show the variation of the square of the speed of sound as a function of baryon number density for different temperatures. The black dot-dashed (green dashed) curve represents the variation of the square of the speed of sound at temperature $T = 10\ (20)\ {\rm MeV}$ while the solid red curve represents the same in the zero-temperature limit. As may be noted, the speed of sound is higher at high temperatures in low-density regions and remains flatter, while it remains the same across different temperatures in higher densities. This is a manifestation of stiffening of the \ac{eos} at lower densities with temperature. At high densities, the speed of sound is almost the same as that of the zero-temperature as the \ac{eos} does not change with temperature as is observed in FIG \ref{fig:equation_of_state}. It may be noted that the speed of sound is an important parameter to estimating the $f$-mode frequencies of \ac{ns} as we shall see later. 

\begin{figure*}[htb]
    \centering
    \includegraphics[scale=0.45]{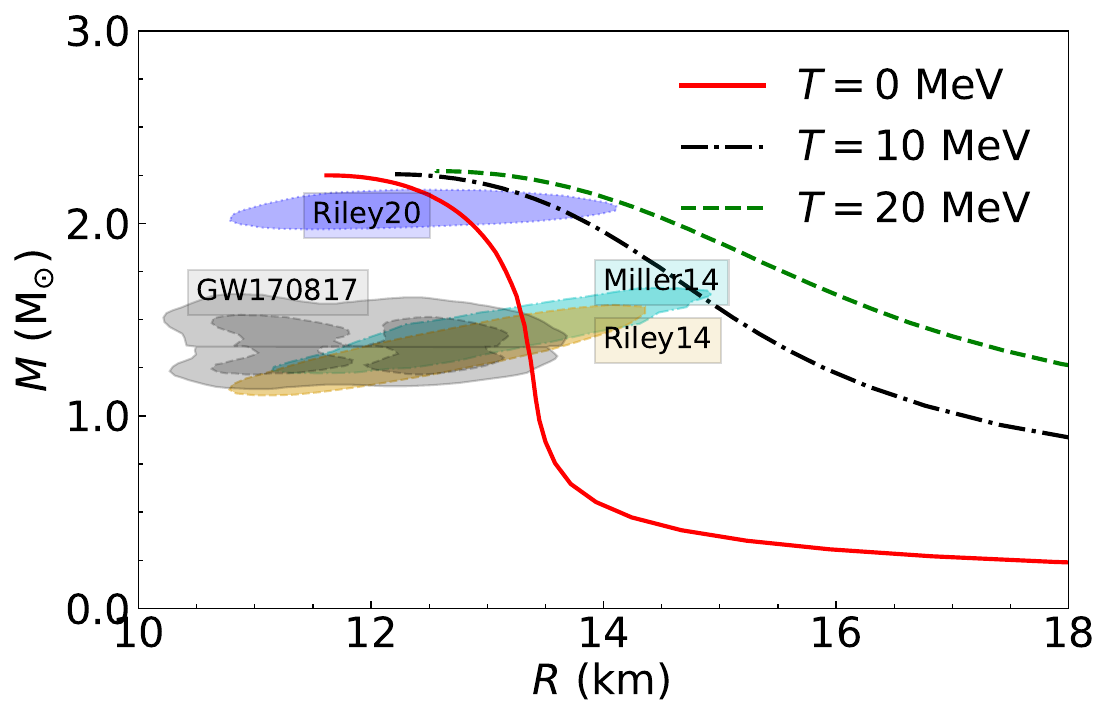}
    \includegraphics[scale=0.45]{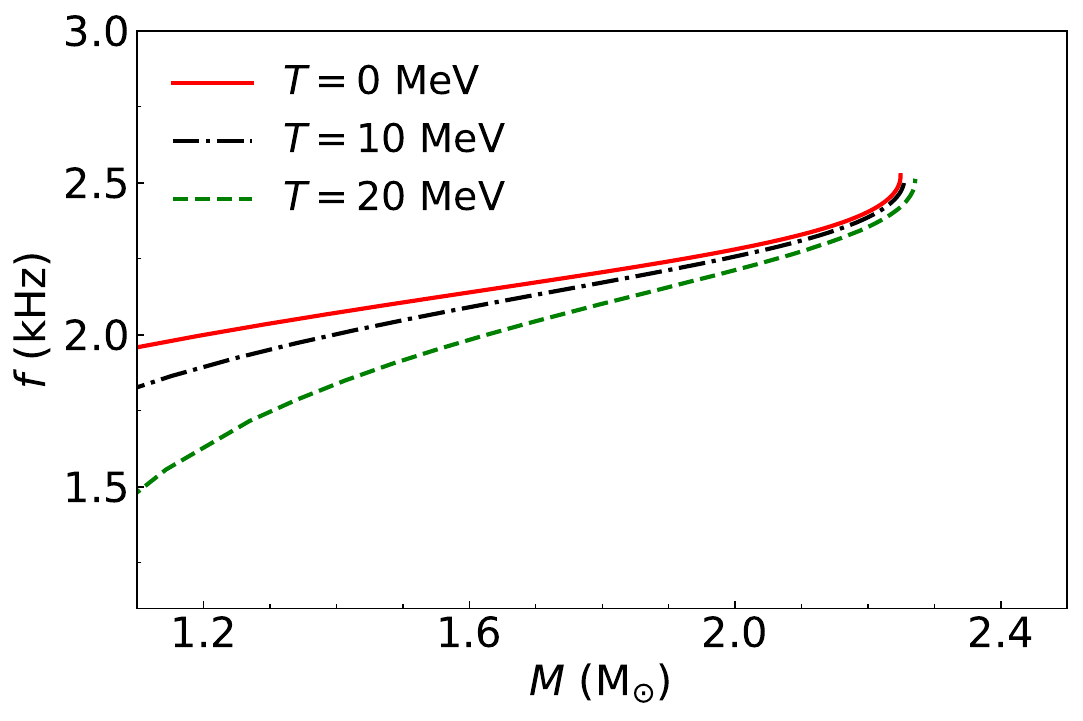}
    \caption{\blue The mass-radius relationships for various $\beta$-equilibrated \ac{eos} depicted in Fig. \ref{fig:equation_of_state} are illustrated on the left panel along with the various astrophysical observations. The gray region corresponds to the GW170817 observation \cite{LIGOScientific:2018hze}. The dark gray and light gray regions here correspond to the 50\% and 90\% confidence interval respectively. The blue patch with dotted outline is the NICER observation of pulsar PSR J0740+6620 of mass $2.14^{+0.1}_{-0.09}\ {\rm M}_{\odot}$ \cite{Riley:2021pdl} while the yellow and cyan with dashed outlines are the NICER observations of a pulsar PSR J0030+0451 \cite{Riley:2019yda, Miller:2019cac} of mass about $1.4\ {\rm M}_{\odot}$. The right panel depicts the nonradial oscillation $f$-mode frequencies, $f=\omega/{2\pi}$, (linear frequencies $f$) as a function of mass for different \ac{ns} at various temperatures (i.e. for the proto-\ac{ns}s), corresponding to the \ac{rmf} model EOS (illustrated in Fig. \ref{fig:equation_of_state}).}
    \label{fig:mass_radius_curve_combined}
\end{figure*}
We have estimated the \ac{eos} at various temperatures, $T=0,\ 10,\ 20\ {\rm MeV}$. We next present the M-R relations of isothermal \ac{ns}s at temperatures $T=0,\ 10,\ 20\ {\rm MeV}$, respectively, by solving the \ac{tov} equations, as given in Eqs. (\ref{tov.pressure}) and (\ref{tov.mass}). {\blue In Fig. \ref{fig:mass_radius_curve_combined} (left), we plot the M-R curves corresponding the \ac{eos} of \ac{nsm} at various temperatures. Here, the red solid curve represents the M-R curve for cold \ac{ns}s while black dot-dashed curve (green dashed curve) represents the same for hot \ac{ns} i.e., proto-neutron star for temperature $T=10\ (20)\ {\rm MeV}$. We show in the same figure the constraints the various astrophysical observations, which constrain the \ac{eos}. The highest observed mass to date is \( 2.14^{+0.1}_{-0.09} \ {\rm M}_{\odot} \) at a 68\% confidence interval for the pulsar PSR J0740+6620, represented by the violet band in the \ac{nicer} x-ray data \cite{Riley:2021pdl}. For completeness, we also present the Bayesian parameter estimation of the mass and equatorial radius of the millisecond pulsar PSR J0030+0451, as reported by the \ac{nicer} mission. The inferred \( M, R \) values from the collected data, shown as cyan and yellow regions, are \( 1.36^{+0.15}_{-0.16} \ {\rm M}_{\odot} \) and \( 12.71_{-1.19}^{+1.14} \) km \cite{Riley:2019yda} and \( 1.44^{+0.15}_{-0.14} \ {\rm M}_{\odot} \) and \( 13.02^{+1.24}_{-1.06} \) km \cite{Miller:2019cac}, respectively. In addition to \ac{nicer} data, we also include constraints derived from LIGO/Virgo gravitational wave observations of GW170817. The gray regions at the top and bottom indicate the 90\% (solid) and 50\% (dashed) confidence intervals from the LIGO/Virgo analysis for each binary component of the GW170817 event \cite{LIGOScientific:2018hze}. As may be observed from the figure, $T=0\ {\rm MeV}$ M-R curve satisfies all the observational constraints. On the other hand at finite-temperature, because of thermal contributions to the pressure compare the $T=0\ {\rm MeV}$ case, one has to integrate out to a larger distance while solving the \ac{tov} equations. This results in larger radii for the hotter \ac{ns}s. As the star cools down, it becomes more compact because the pressure from the thermal part of the \ac{eos} decreases.}

In Fig. \ref{fig:mass_radius_curve_combined} (right) we show the nonradial $f$-mode oscillation frequencies of different mass of isothermal \ac{ns}s corresponding to different temperatures, $T = 0,10,20$ MeV. The red solid curve corresponds to the cold (zero temperature) \ac{ns}s while the black dot-dashed (green dashed curve) corresponds to the hot \ac{ns}s of temperature $T=10\ (20)$ MeV. As observed, the $f$-mode oscillation frequency decreases with the increasing temperature for a given mass of a \ac{ns}. The $f$-mode frequency is inversely proportional to the speed of sound as may be noted in Eq. (\ref{qprime}). Thus the decrease of the $f$-mode frequency with temperature is due to the associated increase of speed of sound with temperature as is observed in Fig. \ref{fig:speed_of_sound_and_particle_densities} (right). It may however be noted that such a conclusion is based on the results within the Cowling approximation which itself is overestimated with respect to a frequency estimation in the general relativistic approximation \cite{Yoshida:1997bf, Kumar:2024abb}. However, the qualitative behavior of decrease of $f$-mode frequency at finite-temperature can remain still valid while the quantitative value may differ. The $f$-mode oscillation has no node inside the star and is directly correlated with the mass-radius relation. The $f$-mode frequency differs considerably for low-mass stars; however, it is very similar for massive stars, irrespective of the temperature. Therefore, for relatively low-mass stars, the detection probability increases with temperature. From this study, it is clear that the observation of nonradial oscillation frequency of hotter \ac{ns} is less challenging since for the same mass \ac{ns}s, the $f$-mode oscillation frequency is smaller for hotter \ac{ns} as compared with colder one. Similar results for the nonradial oscillation frequencies of proto-neutron stars have been found in recent studies \cite{Lozano:2022qsm}. The frequency range for the \ac{eos}s with different temperatures is consistent with the known range of $f$-mode frequencies ($\sim$ 1 to 3 kHz).

In the previous studies \cite{Kumar:2023rut, Kumar:2023ojk, Sotani:2023zkk, Li:2023owg, Ranea-Sandoval:2023ixr, Andersson:1997rn, Benhar:2004xg}, various \ac{ur}s relating the oscillation frequency, mass, radius, moment of inertia, tidal deformability and other \ac{ns} properties have been proposed for the cold \ac{ns}s, which are very important to reduce the dependence on \ac{eos}. Further, such relations help to estimate the physical variables which are difficult to measure in terms of variable which are easier to measure in astrophysical observations. In the current study, we attempt to find such a \ac{ur} among $f$-mode frequency, mass and radius for the isothermal \ac{ns}s. 

\begin{figure}[htb]
    \centering
    \includegraphics[scale=0.45]{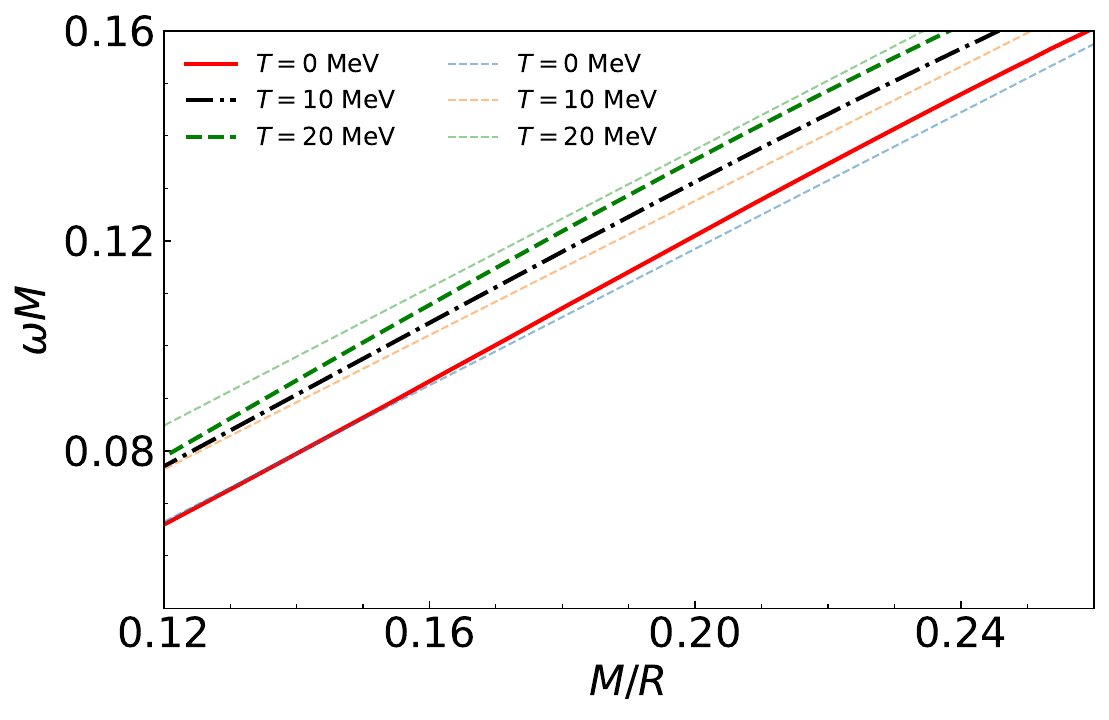}
    \caption{\blue The variation of dimensionless quantity $\omega M$ as a function of dimensionless quantity ($M/R$), the compactness of NSs. The red solid thick lines indicate the same variation for zero-temperature NS while black dot-dashed thick (green dashed thick) line shows the same for isothermal proto-neutron stars of temperature ($T=0,10,20$ MeV). The thin lines correspond to the universal relations for the $f$-mode frequencies as in Eq. (\ref{eq:universal_relation}) for proto-neutron stars at different temperatures ($T=0,10,20$ MeV) using eight \ac{eos}s from which the seven the EOSs are taken from Table \ref{tab:eos_models} along with the present RMF EOS.
    }
    \label{fig:universal_relations_with_temp_combined}
\end{figure}

{\blue In Fig. \ref{fig:universal_relations_with_temp_combined}, we plot the angular $f$-mode frequency times the mass of \ac{ns}, $\omega M$, as a function of compactness, $C=M/R$, for the present \ac{rmf} model shown by the red solid line for $T=0\ {\rm MeV}$, black dash-dotted line for $T=10\ {\rm MeV}$, and the green dashed line for $T=20\ {\rm MeV}$. To examine a universal behavior relating these two variables, we have considered seven more \ac{eos}s listed in Table \ref{tab:eos_models} which have been taken from the CompOSE repository. It may be mentioned that all these seven \ac{eos}s satisfy all the constraints from the astrophysical observations which are shown in Fig. \ref{fig:mass_radius_curve_combined} (left). We estimated the $f$-mode frequencies of \ac{ns}s corresponding to each of these seven \ac{eos}s along with their masses and radii. Using these seven EOSs along with the present RMF EOSs, it turns out that the variables e.g., $\omega M$ and $M/R$ can be fitted with the \ac{ur} $\omega M = a \left(\frac{M}{R}\right) + b$ for various isothermal \ac{ns}s. For a given temperature, the coefficients of \ac{ur}, $a(T)$ and $b(T)$ are estimated so as to satisfy the \ac{ur} for $\omega M$ and $M/R$ values of the total eight \ac{eos}s at that temperature. We have plotted the best fitted curves, the \ac{ur}s, in Fig. \ref{fig:universal_relations_with_temp_combined} as corresponding thin lines for different temperatures. It may be noted that there are different \ac{ur}s considered in the literature relating $\omega$, $M$ and $R$ e.g $\omega \propto \sqrt{\frac{M}{R^3}}$ \cite{Benhar:2004xg, Andersson:1997rn}, but these relations are not found universal for the isothermal \ac{ns}s. In this context, we had earlier considered an \ac{ur} using a very large set of \ac{eos}s relating these variables for cold \ac{ns} as above which turned out to be quite robust \cite{Kumar:2023rut}. As may be observed in Fig. \ref{fig:universal_relations_with_temp_combined} that the \ac{ur} for zero temperature is consistent with the results discussed previously \cite{Kumar:2023rut}. As temperature increases, the \ac{ur} shifts towards larger $\omega M$ for given compactness; however, the shift is not linear}. To understand this behavior we made the coefficients in the \ac{ur} $a$ and $b$ temperature dependent keeping up to quadratic dependence on temperature. Thus the \ac{ur} can be recast as
\begin{IEEEeqnarray}{rCl}
\omega M &=& a(T) \left(\frac{M}{R}\right) + b(T),
\label{eq:universal_relation}
\end{IEEEeqnarray}
where $a(T)$ and $b(T)$ are defined as 
\begin{IEEEeqnarray}{rCl}
a(T) &=& a_0 + a_1 T + a_2 T^2, \label{eq:a}\\
b(T) &=& b_0 + b_1 T + b_2 T^2. \label{eq:b}
\end{IEEEeqnarray}

\begin{table}[h]
\caption{The coefficients of \ac{ur}s, Eqs. (\ref{eq:a}) and (\ref{eq:b}).}
\begin{tabular}{lr}
\toprule
Parameters & Numerical Value \\
\hline
$a_0$ & $0.65257 \pm 8.7\times 10^{-6}$ \\
$a_1$ & $-0.00284 \pm 0.00018\ {\rm MeV}^{-1}$ \\
$a_2$ & $0.00016 \pm 0.00078\ {\rm MeV}^{-2}$ \\
$b_0$ & $-0.01266 \pm 4.09\times 10^{-6}$ \\
$b_1$ & $0.00155 \pm 8.47\times 10^{-5}\ {\rm MeV}^{-1}$ \\
$b_2$ & $-0.00003 \pm 0.00037\ {\rm MeV}^{-2}$ \\
\hline
\end{tabular}
\label{tab:ur_parameters}%
\end{table}

\begin{figure*}[htb]
    \includegraphics[scale=0.40]{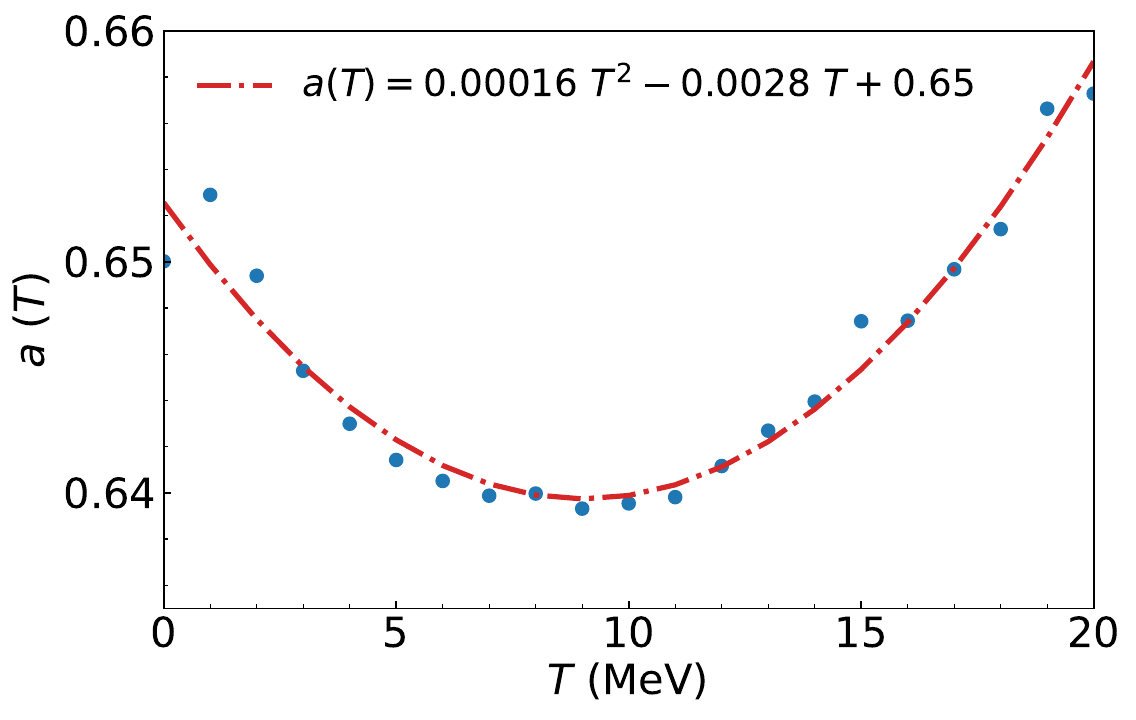}
    \includegraphics[scale=0.40]{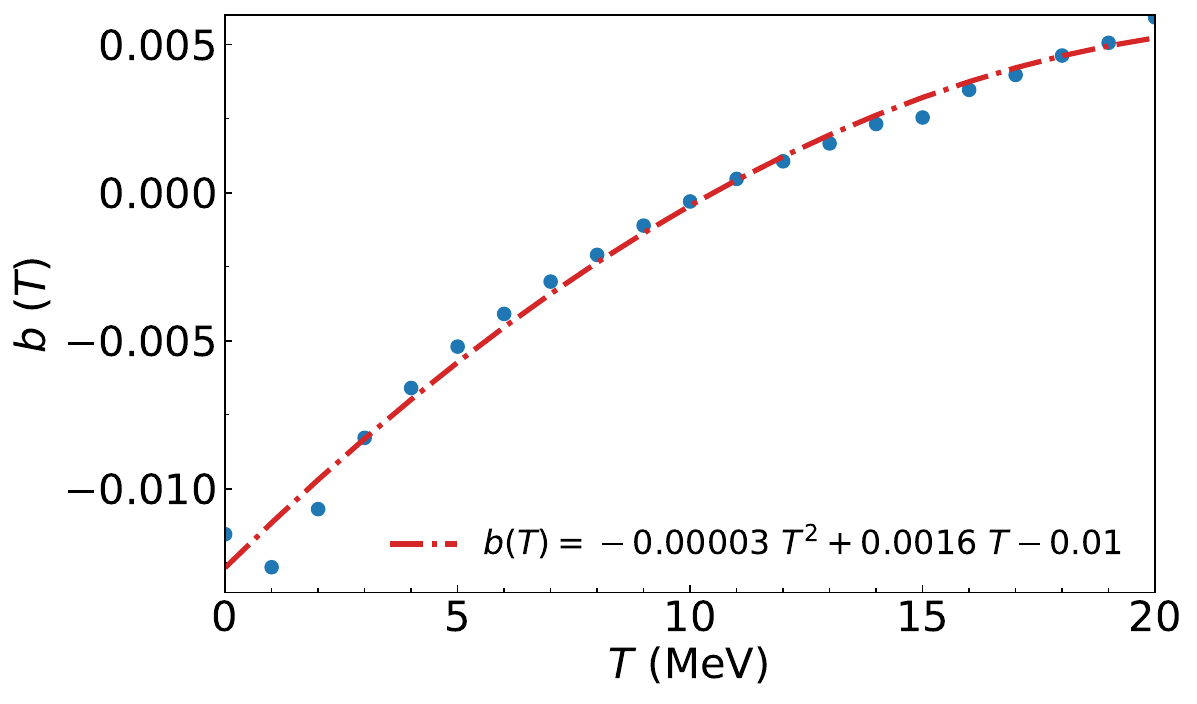}
    \caption{The slope, $a(T)$, (left) and intercept, $b(T)$, (right) of the linear fit between $\omega M$ and $\frac{M}{R}$ as a function of temperatures.}
    \label{fig:tm_and_tc}
\end{figure*}
The values of the parameters are given in Table \ref{tab:ur_parameters}. It is instructive to check how the parameters vary with temperature. Therefore, we plot the temperature dependence of the parameters in Fig. \ref{fig:tm_and_tc}. In Fig. \ref{fig:tm_and_tc} (left), it may be observed that the slope parameter of the \ac{ur}, $a(T)$, has a parabolic dependence with temperature with a minimum at $T\sim 9\ {\rm MeV}$. On the other hand, the intercept parameter of \ac{ur}, $b(T)$, shows a monotonic increase with temperature.

\section{Summary and conclusion} \label{summary_and_conclusion}
In this study, we investigate the impact of temperature on various properties of \ac{ns}s, specifically focusing on the M-R relation, the fundamental-mode ($f$-mode) frequency, and \ac{ur}s. The temperature influences the \ac{eos} of the star, which in turn affects these NS properties. Our analysis considers simple $npe$ nuclear matter, and we have developed an \ac{rmf} model where the couplings are density dependent, consistent with nuclear saturation properties. The set of equations used ensures that the resulting neutron stars adhere to astrophysical observations for cold \ac{ns}s. It is also assumed that temperature does not directly affect the oscillation equations in the sense that the oscillation equations do not depend explicitly on temperature and the temperature dependence arises implicitly through the temperature dependence of \ac{eos}.

{\blue Our findings also indicate that temperature predominantly affects the low-density regions of the matter, with minimal impact at higher densities as the temperatures considered are much smaller compared to the chemical potentials. This results in a negligible variation with temperature. This is consistent with earlier observations \cite{Lattimer:1991nc, Kochankovski:2022rid, Tsiopelas:2024ksy}. On the other hand, as temperature increases, low-density matter becomes stiffer, leading to a flatter mass-radius curve (i.e., larger radii for the same mass \ac{ns}). Consequently, in mass-radius sequences derived from temperature-dependent \ac{eos}s, low-mass stars, which correspond to lower central energy densities, exhibit significantly larger radii as the temperature rises.}

The alterations in the \ac{eos} and the corresponding changes in the speed of sound significantly influence the $f$-mode frequency, causing it to shift towards lower values with increasing temperature. This suggests that $f$-mode frequencies for proto-neutron stars are more easily detecTable with gravitational wave detectors. Additionally, we have explored the generalized \ac{ur}s in terms of nonradial oscillation $f$-mode frequency, radius, and mass of neutron stars. The \ac{ur} relation demonstrates a nonlinear behavior with temperature, and when parameterized, the parameters exhibit an oscillatory pattern. This oscillation is primarily due to the temperature effect on the \ac{eos}, with low-density matter being more affected than high-density matter, causing the lower region of the universal relation to align across various temperatures. This nonlinearity is also evident in the mass-radius sequence, where for the same mass, a higher radius is observed (corresponding to lower energy densities) as temperature increases.

\section{Acknowledgment}
D.K. and R.M. would like to acknowledge the financial support from the Science and Engineering Research Board (SERB), Govt. of India, grant ``Core Research Grant (CRG/2022/000663)". A.V. acknowledges the Prime Minister’s Research Fellowship (PMRF), Ministry of Education Govt. of India, for a graduate fellowship.

\bibliographystyle{ieeetr}
\bibliography{bib}

\end{document}